\documentclass[reprint,aps,prl,showpacs,twocolumn, superscriptaddress]{revtex4-2}
\usepackage{graphicx}
\usepackage{dcolumn}
\usepackage{bm}
\usepackage{amsmath}
\usepackage{amssymb}
\usepackage{physics}
\usepackage{lipsum}
\usepackage{graphics}
\usepackage{xcolor}
\usepackage{bbold}
\usepackage{mathtools}
\usepackage{natbib}
\usepackage{hyperref}
\usepackage{bookmark}
\hypersetup{
    colorlinks,
    linkcolor={blue},
    citecolor={blue},
    urlcolor={blue}
}

\usepackage{multirow}
\usepackage{bbm}
\usepackage{xr}
\usepackage{amsthm, amssymb, mathrsfs}
\usepackage{empheq}
\usepackage{cases}

\begin{document}

\title{Universal Statistics of Competition in Democratic Elections}

\author{Ritam Pal}
\email{ritam.pal@students.iiserpune.ac.in}
\affiliation{Department of Physics, Indian Institute of Science Education and Research, Pune 411008, India.}
\author{Aanjaneya Kumar}
\altaffiliation[Present address: ]{Santa Fe Institute, 1399 Hyde Park Road, Santa Fe, NM 87501, USA}
\email{aanjaneya@santafe.edu}
\affiliation{Department of Physics, Indian Institute of Science Education and Research, Pune 411008, India.}
\author{M. S. Santhanam}
\email{santh@iiserpune.ac.in}
\affiliation{Department of Physics, Indian Institute of Science Education and Research, Pune 411008, India.}

\date{\today}

\begin{abstract}
Elections for public offices in democratic nations are large-scale examples of collective decision-making. As a complex system with a multitude of interactions among agents, we can anticipate that universal macroscopic patterns could emerge independent of microscopic details. Despite the availability of empirical election data, such universality, valid at all scales, countries, and elections, has not yet been observed. In this work, we propose a parameter-free voting model and analytically show that the distribution of the victory margin is driven by that of the voter turnout, and a scaled measure depending on margin and turnout leads to a robust universality. This is demonstrated using empirical election data from $34$ countries, spanning multiple decades and electoral scales. The deviations from the model predictions and universality indicate possible electoral malpractices. We argue that this universality is a stylized fact indicating the competitive nature of electoral outcomes.
\end{abstract}

\maketitle

One of the cornerstones of democratic societies is that governance must be based on an expression of the collective will of the citizens. The institution of elections is central to the operational success of this system. Elections to public offices are the best-documented instances of collective decision-making by humans, whose outcome is determined by multiple agents interacting over a range of spatial and temporal scales. These features make elections an interesting test-bed for statistical physics whose key lesson is that a multitude of complex interactions between microscopic units of a system can manifest into robust, {\it universal} behavior at a macroscopic level \cite{anderson1972more,strogatz2022fifty,CasForLor2009, JedSzn2019, MigTor2020, galam2012, brams2008, ForMacRed2013, Bouchaud2023, SenCha2014, PerJorRan2017,JusHolKan2022,redner2019reality}. A collection of gas molecules or spins are examples that display such emergent macroscopic features \cite{REI65}, and so are complex processes such as earthquakes \cite{Corral2004,Corral2006} and financial markets \cite{PleGopRos1999}. In the context of elections, such universal behaviors serve to distill the complexities of electoral dynamics into understandable and predictive frameworks and safeguard its integrity. 

Unsurprisingly, the possibility of universality in elections attracts significant research attention \cite{CosAlmAnd1999, ForCas2007, BorBou2010, mantovani2011scaling, BokSzaVat2018, ChaMitFor2013, hosel2019universality}. Several works have studied and proposed models for ({\it a}) the distribution $q(\sigma)$ of the fraction of votes $\sigma$ obtained by candidates (or the vote share), and ({\it b}) distribution $g(\tau)$ of voter turnout $\tau$. While $\sigma$ is indicative of popularity, $\tau$ indicates the scale of the election. Though some universality has been observed in $q(\sigma)$ or $g(\tau)$ within a single country \cite{ForCas2007,CosAlmAnd1999, BorBou2010} or in countries with similar election protocols \cite{ForCas2007, ChaMitFor2013}, deviations from claimed universalities have also been reported \cite{ChaMitFor2013, Kon2017,Kon2019, CalCroAnt2015, BorRayBou2012} due to variations in the size (scale) of electoral districts and weak party associations. Though voting patterns tend to display spatial correlations \cite{FerSucRam2014, BraDeA2017,MicIlkAtt2021,MorHisNak2019}, it is not known to be universal. Despite the availability of enormous election data and persistent attempts, a robust and universal emergent behavior, valid across different scales and countries with vastly different election protocols, is yet to be demonstrated.

In this Letter, using extensive election data \cite{india_data, clea, canada_data, DVN/VOQCHQ_2018} from 34 countries (from 6 continents) spanning multiple decades and electorate scales, we demonstrate universality through analysis of the margin of victory and turnout data in democratic elections. The \emph{margin of victory} (or simply the \emph{margin}) is a key indicator of competition in elections and a proxy for the healthy functioning of democracies. While the turnout data has been studied in various settings, margins have never been considered in the context of universality. We propose a Random Voting Model (RVM) and demonstrate that the turnout distribution drives the distribution of scaled margin, {\it i.e.} the model predicts the scaled margin distribution with only the turnout distribution as the input. We analytically derive the distribution of scaled margin-to-turnout ratio in the RVM and show that it exhibits universal characteristics independent of the turnout distribution. Remarkably, we find that empirical election data across 32 countries shows excellent agreement with the analytical results, establishing a robust universality. We demonstrate its utility as a novel statistical indicator for flagging electoral malpractices \cite{klimek2012statistical, jimenez2017testing}.

\begin{figure*}[t!]
    \centering
    \includegraphics[width=\textwidth]{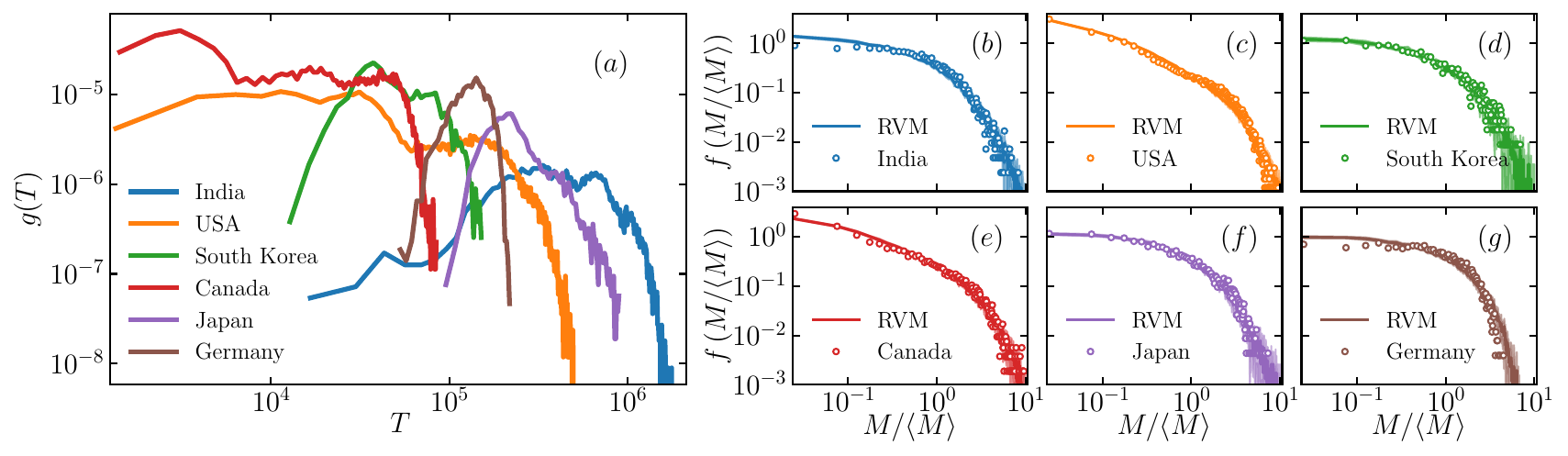}
    \caption{(a) Turnout distribution $g(T)$ obtained from election data for different countries. Note the differences in shapes and ranges for $g(T)$. (b-g) Scaled margin distribution $f(M/\langle M \rangle)$ obtained from election data (open circles) and the model predictions (solid lines) display an excellent agreement. The lighter shade around the model prediction represents its variability estimated from multiple RVM realizations.}
    \label{fig_1}
\end{figure*}
A template of a basic electoral process is as follows. At each electoral unit, candidates compete against each other to win the votes of the electorate, who can cast their vote in favor of only one of the candidates. The candidate securing the largest number of polled votes is declared the winner. This represents the core process in many electoral systems. It is the standard first-past-the-post system followed in many countries, e.g., India, the UK, and the USA. In an instant-run-off system (such as in Australia) or two-round run-offs (such as in France), the final run-off round boils down to this template. Typically, national or regional elections following this template consist of many electoral units made up of polling booths, precincts, constituencies, or counties. These units set a size scale in terms of the number of electorates -- polling booth represents the smallest scale, while a constituency (subsuming many polling booths) represents the largest scale. For our analysis, an ``election'' could be either a national, regional, or even a city-level electoral process encompassing $N$ electoral units, and each unit could be a polling booth, county, or constituency.

In any such election, an informative indicator of the degree of competition and the extent of consensus is the margin. A vanishing margin signifies tight competition and a divided electorate, whereas large margins indicate a decisive mandate and overwhelming consensus in favor of one candidate. Let $c_i, i=1, 2, \dots N$, denote the number of candidates contesting an election in the $i$-th electoral unit. The winning and runner-up candidates receive, respectively, $v_{i, w}$ and $v_{i, r}$ votes such that $v_{i, w} > v_{i, r}$. The margin is given by $M_i=v_{i, w}-v_{i, r}$. If $n_i > 0$ is the size of the electorate, {\it i.e.}, number of registered voters in $i$-th unit, then $0 \le M_i \le n_i$. However, in practice, only a fraction of the electorate participates in voting. In such cases, the number of voters who show up to cast their vote is termed as the turnout $T_i$, such that $0 \le T_i \le n_i$, and consequently, the margin is further restricted by $0 \le M_i \le T_i$.

\begin{figure*}[t]
    \centering
    \includegraphics[width=\textwidth]{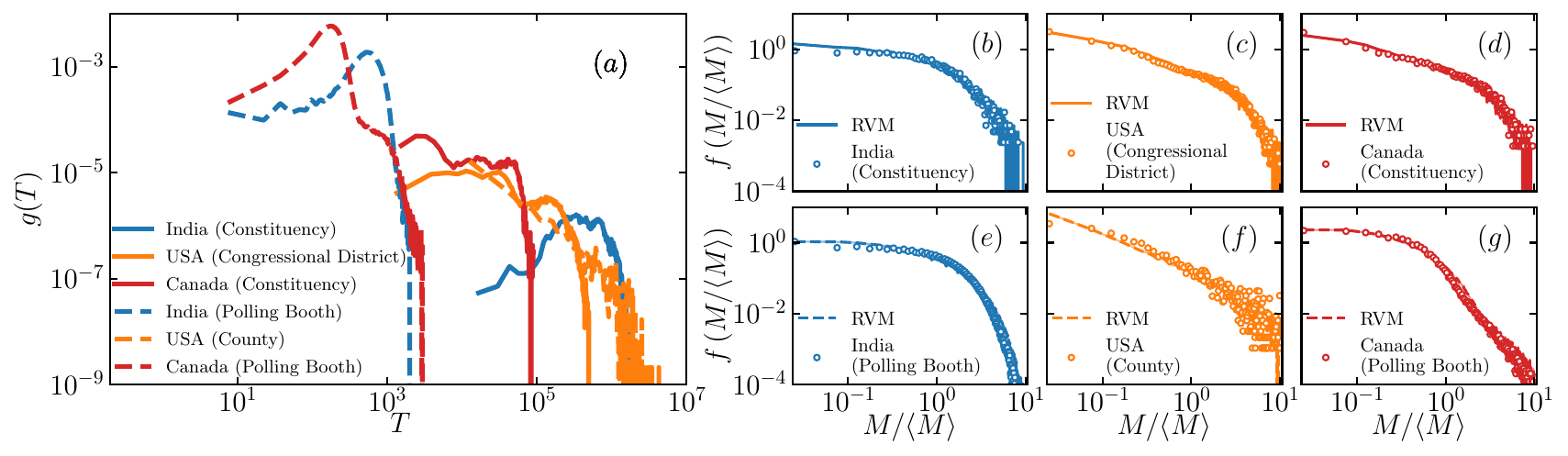}
    \caption{The turnout distribution $g(T)$ and scaled margin distribution $f(M/\langle M \rangle)$ for India (blue), the USA (orange), and Canada (red), at two widely different scales, {\it i.e.}, size of electoral units. (a) $g(T)$ at two different scales for each country. The dashed line is for smaller scales (polling booth for India and Canada, County for the USA), while the solid line represents a larger scale (constituency for India and Canada, congressional district for the USA). (b-g) $f(M/\langle M \rangle)$ from election data (open circles) and as predicted by the RVM (line). Despite the differences in scale and shape of $g(T)$, the empirical $f(M/\langle M \rangle)$ is well described by the RVM. The lighter shade around the model prediction represents variability estimated from multiple RVM realizations.}
    \label{fig_2}
\end{figure*}

To fix our ideas, we might focus on the elections in one country, e.g., the general elections in India. Then, the object of interest would be $M_i$ and $T_i$ ($i=1,2, \dots N$). To be statistically robust, the data is consolidated from many elections spread over several decades (For India, $18$ elections from 1951 to 2019; See Sec.~S6 of supplemental Material \cite{supp})\nocite{abramowitz_stegun}. This leads to the associated empirical distributions $Q(M)$ and $g(T)$, respectively, for margin and turnout. Figure \ref{fig_1}(a) displays the distribution of raw turnout $g(T)$ at the constituency level for national elections in six countries, namely, India, USA, South Korea, Canada, Japan, and Germany. Striking dissimilarities in $g(T)$ are visible in the shape and support of distribution for countries. For Germany, $g(T)$ has a unimodal character, while that for Canada and the USA display multiple peaks. The corresponding scaled margin $M/\langle M \rangle$ is displayed as distribution $f(M/\langle M \rangle)$ (computed from the consolidated margin data for each country) in Fig.~\ref{fig_1}(b-g). While they appear to be broadly similar, certain differences are clearly noticeable. In particular, $f(M/\langle M \rangle)$ for German elections in Fig.~\ref{fig_1}(g) has a sharp cutoff, but for India and Japan in Fig.~\ref{fig_1}(b, f) the distribution has a slower decay. These observations motivate the questions of whether $f(M/\langle M \rangle)$ is related to the raw turnout distribution and can be obtained from it.

To investigate this question, we propose a Random Voting Model (RVM) ${\mathcal V}(T)$ that takes raw turnouts $T=\{T_1, T_2 \dots T_N\}$ as input. This model emulates an election taking place at $N$ electoral units (say, constituencies). At $i$-th unit, each of the $T_i$ voters (raw turnout at $i$-th unit) can cast only one vote, independently and by randomly choosing one of the $c_i$ contesting candidates. The probability that candidate $j$ in $i$-th unit can attract a vote is $p_{ij}= w_{ij}/\sum_k w_{ik}$, where $w_{ij} \in [0,1]$ is a random number drawn from a uniform distribution. While this protocol provides a natural and effective choice for $p_{ij}$, the sensitivity of the RVM predictions on different protocols is discussed in Sec.~S5 of Ref. \cite{supp}. In election data that we use, averaged over all the 34 countries, the top two (three) candidates account for 79\% (87\%) of all votes polled. Hence, the model assumes three candidates at every constituency: $c_i=3$ for $i=1,2 \dots N$, and that all eligible voters cast their votes, implying $T_i = n_i$. By simulating this model, margin $M_i$ is obtained for $i$-th electoral unit and $\langle M \rangle = (1/N) \sum_{i=1}^N M_i$ is the associated sample mean. For a detailed description of the model, see Sec.~S1 of Ref. \cite{supp}. 

The model predictions depend exclusively on the actual turnout distribution, and no free parameters to be tuned. As illustrated in Fig.~\ref{fig_1}(b - g), the scaled margin distributions predicted by this model (solid lines) show a remarkable agreement with those computed from empirical margin data from real elections. Notably, RVM faithfully captures disparate decay features in  $f(M/\langle M \rangle)$ for India, USA, South Korea, Canada, Japan, and Germany (for 28 other countries, see Sec.~S7 of Ref. \cite{supp}). This suggests that the raw turnout data carries intrinsic information about the margin distribution. RVM effectively leverages this information embedded in the turnout distribution to predict the scaled margin distribution. 

\begin{figure*}[t]
    \centering
    \includegraphics[width=\textwidth]{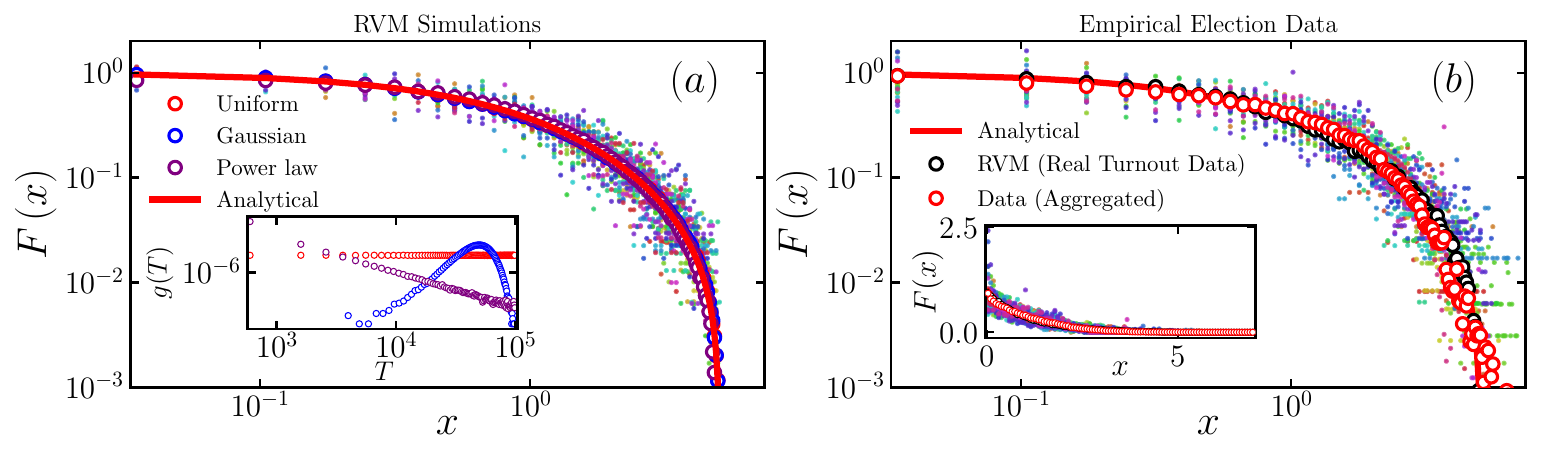}
    \caption{(a) $F(x)$ predicted by RVM for three different turnout distributions $g(T)$ (see inset). The open circles are obtained from RVM simulations with $N = 10^6$, while the solid colored circles are generated from RVM simulation with $N$ identical to empirical election data. The red line corresponds to $F(x)$ in Eq. \ref{eq:Fscale}. (b) The empirical distribution of $x=\mu/\langle \mu \rangle$ from election data of 32 countries (excluding Ethiopia and Belarus). Each color indicates a specific country for which the empirical election data is consolidated over several elections. The average of these empirical distributions (red open circles) closely follows the analytical curve (red line) and the averaged RVM predictions for each country (black open circles). The inset depicts the distributions on a linear scale.}
    \label{fig_3}
\end{figure*}

Next, we show that these results are independent of the number of voters or size of electoral units. In large countries, depending on the size of the electoral unit, the typical turnout can differ by several orders of magnitude. For example, in India, polling booths have a typical electoral size $\sim 10^3$, whereas, at the parliamentary constituency level, it is about $10^6$. Further, the shapes of $g(T)$ are also vastly different at different scales. Figure \ref{fig_2}$(a)$ captures the striking differences in range and shape of $g(T)$ for India, the US, and Canada at two different scales. Quite remarkably, despite these vast differences in the scale, the same RVM ${\mathcal V}(T)$, without any parameter adjustments, accurately predicts the scaled margin distribution. Figure \ref{fig_2}$(b, c, d)$ shows the empirical distribution of scaled margins (in national elections) at the constituency-level scale, and Figure \ref{fig_2}$(e, f, g)$ shows the same at the scale of polling booths (county for USA). The margin distribution computed from the model is in agreement with the empirical distribution at both scales. Theoretical analysis in the limit $T \gg 1$ (see Sec.~S3 of Ref. \cite{supp}) shows that the tail of $g(T)$ dictates the tail of the $f(M / \langle M \rangle)$. This is confirmed by the RVM simulations (see Sec.~S4 of Ref. \cite{supp}). In particular, this is evident for the USA, where the county-level turnout distribution shows a heavy-tailed decay, which is reflected in the corresponding scaled margin distribution (Fig.~\ref{fig_2}(f)). The faster decay at congressional district level distribution (Fig.~\ref{fig_2}(c)) is also predicted by RVM. For Canada too, the empirical scaled margin distributions are noticeably different at two different scales. Yet, the differences are well captured by the RVM simulations shown as dashed and solid lines in Fig.~\ref{fig_2}(b-g). Taken together, these results show that the scaled margin distribution depends on the raw turnout distribution, and RVM captures this relation across various countries and at all scales. Then, a relevant quantity of interest would be the ratio $\mu = \frac{M}{T}$, to be called the specific margin, with  $0<\mu<1$. This is a turnout-independent measure of electoral competitiveness and does not depend on the size of the electorate. 

To obtain analytical insight, we consider elections with three candidates in the limit of large turnout ($T \gg 1$). The votes received by $j$-th candidate can be approximated as $v_j \approx p_jT$, and the margin as $M \approx \left(p_{(3)} - p_{(2)}\right) T$, where $p_{(k)}$ denotes $k$-th order statistics \cite{BarBalNag2008} of the probabilities assigned to the candidates. Evidently, in this limit, $\mu \approx p_{(3)} - p_{(2)}$ and its distribution has no explicit dependence on $T$. With this insight, we obtain the distribution of specific margins as \cite{supp}

\begin{equation}
    P\left(\mu\right) =  \frac{(1 - \mu)(5 + 7\mu)}{(1 + \mu)^2(1 + 2\mu)^2}. 
\end{equation}
Thus, the distribution $F(x)$ of the scaled specific margin $x = \mu / \langle \mu \rangle$, can be expressed as
\begin{equation}
 F\left(x\right) = \langle \mu \rangle ~ P\left( x \langle \mu \rangle \right),   
\label{eq:Fscale}
\end{equation}
with $\langle \mu \rangle = \frac{1}{2}+\ln \left(\frac{9 \sqrt[4]{3}}{16}\right)$. 
Figure \ref{fig_3}(a) demonstrates that $F(x)$, computed from RVM simulations with vastly different turnout distributions $g(T)$, does not depend on the detailed structure of $g(T)$ and is in agreement with the analytical prediction in Eq. \ref{eq:Fscale}.

 The RVM simulations are performed with $10^6$ electoral units (for simulation details, see Sec. S4 of Ref.~\cite{supp}) using $g(T)$ corresponding to power law, Gaussian, and uniform distributions (inset of Fig.~\ref{fig_3}(a)). The simulated distributions (open circles in Fig.~\ref{fig_3}(a)), for the three cases of $g(T)$, collapse on the analytical prediction $F(x)$ (red line). 
 
 Bolstered by the ability of RVM to capture the statistics of real elections in Figs.\ref{fig_1}-\ref{fig_2}, we examine if this universality prediction in Eq. \ref{eq:Fscale} holds good for the empirical election data. Indeed, as observed in Fig.~\ref{fig_3}(b), the RVM prediction (black open circles) is in excellent agreement with the averaged distributions (open red circles) obtained from all the $32$ countries. The averaged empirical distribution is also consistent with the analytical universal curve $F(x)$ (red line). Further, the empirical distribution for each of the $32$ countries (denoted by the solid-colored circles) closely follows the trend of $F(x)$, albeit with some fluctuations induced by the finite size of data. Similar fluctuations are evident in RVM simulations as well, seen as solid circles in Fig.~\ref{fig_3}(a), when the number of electoral units $N$ is taken from the empirical election data (rather than fixed at $10^6$) \cite{supp}. Empirical distributions shown in the inset of Fig.~\ref{fig_3}(b) demonstrate that at large $x$, the absolute fluctuations decrease. Thus, the universality in Fig.~\ref{fig_3} suggests that irrespective of the finer details of election processes, the mechanism underlying the core component of any competitive election -- choosing one candidate from many contenders -- leads to a universal distribution for the scaled specific margin $x=\mu/\langle \mu \rangle$.
 
\begin{figure}[t]
    \includegraphics[width=\columnwidth]{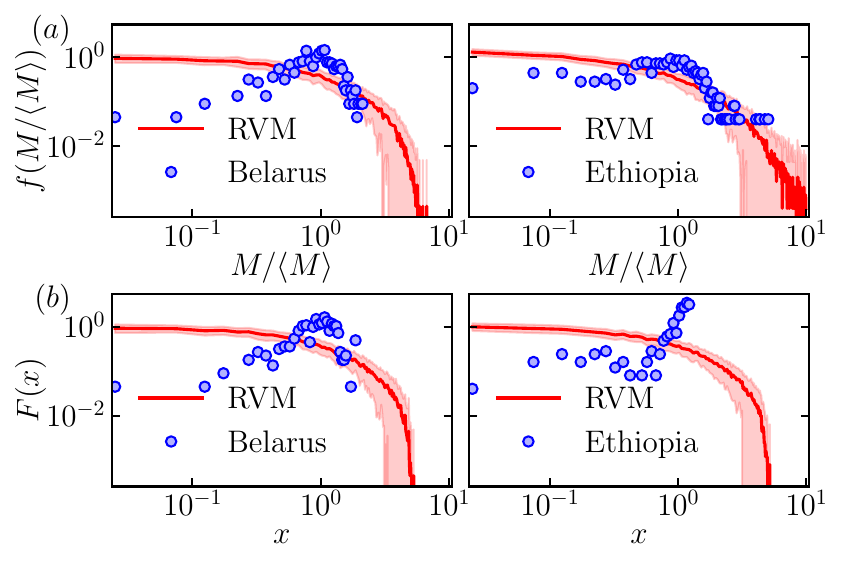}
    \caption{ The distributions $(a)$ $f(M/\langle M \rangle)$, and (b) $F(x)$ obtained from empirical data from Belarus $(2004-2019)$ and Ethiopia $(2010)$ (blue circles). Both show significant deviation from the model predictions (red line). The light red shaded region represents the variability in RVM prediction computed from 100 realizations.}
    \label{fig_4}
\end{figure}
From the excellent RVM predictions of scaled margin distributions (Fig.~\ref{fig_1}, \ref{fig_2}) and the robustness of the universality result (Fig.~\ref{fig_3}) across different countries with a track record of fair election processes, it is reasonable to assume that any pronounced deviation from $F(x)$ in Eq. \ref{eq:Fscale} might indicate a prevalence of unfair means in the election process. We search for such deviations in countries with at least $400$ data points in the constituency-level election data. We find that $F(x)$ computed from data for Ethiopian election of $2010$ and Belarus elections during $2004-2019$ display pronounced deviations from the RVM predictions and universality as seen in Fig.~\ref{fig_4}(b). Similarly, the empirical scaled margin distribution $f(M / \langle M \rangle)$ deviates significantly from the RVM prediction (Fig.~\ref{fig_4}(a)). This analysis in Fig.~\ref{fig_4} strengthens the skepticism expressed in earlier studies and independent investigations about elections in Ethiopia \cite{brigaldino2011elections} and Belarus \cite{belarus_report,frear2014parliamentary,bedford_2021,czwolek2021belarusian}. Electoral malpractices take various forms, and statistical analysis is useful as a prima facie indicator requiring detailed scrutiny. Thus, the robust universality and RVM provide an effective toolbox to flag potentially suspicious elections. We propose that the universality in Fig.~\ref{fig_3} should be treated as a stylized fact of elections, which all election models should be able to reproduce.

In summary, competitiveness in any election is encoded in the victory margins and turnouts. The latter also expresses people's interest in the participatory democratic process. In this work, using extensive empirical election data from $34$ countries, we have obtained two significant results: (a) scaled margin distribution can be predicted from the raw election turnout alone, (b) the scaled distribution of margin-to-turnout ratio $\mu$ has a universal form for all elections independent of country, regions, turnouts and the scale of elections. A parameter-free model introduced in this work faithfully reproduces all these features observed in empirical election data and has been analytically solved to demonstrate universality. Both these results can be regarded as stylized facts of elections. Hence, every successful election model, irrespective of its underlying principle and mechanism, must necessarily reproduce these stylized facts to be consistent with real elections. Further, the deviations from the universal scaling function could potentially help in assessing the credibility of the election process. We demonstrate this by flagging the elections of two countries for possible electoral misconduct.

\begin{acknowledgments}
\emph{Acknowledgements}.---The authors gratefully acknowledge the feedback of an anonymous reviewer whose suggestions greatly improved the manuscript. R.P. and A.K. thank the Prime Minister’s Research Fellowship of the Government of India for financial support.  M.S.S. acknowledges the support of a MATRICS Grant from SERB, Government of India, during the early stages of this work. The authors acknowledge the National Supercomputing Mission for the use of PARAM Brahma at IISER Pune.
\end{acknowledgments}

\newpage
\setcounter{page}{1}
\renewcommand{\thepage}{S\arabic{page}}
\setcounter{equation}{0}
\renewcommand{\theequation}{S\arabic{equation}}
\setcounter{figure}{0}
\renewcommand{\thefigure}{S\arabic{figure}}
\setcounter{section}{0}
\renewcommand{\thesection}{S\arabic{section}}
\setcounter{table}{0}
\renewcommand{\thetable}{S\arabic{table}}

\onecolumngrid
\newpage
\begin{center}
\textbf{\large Supplemental Material for ``Universal Statistics of Competition in Democratic Elections"}
\end{center}

This Supplemental Material provides further discussion and derivations which support the findings reported in the Letter, and provides details of the models and simulations used to validate the results. 

\tableofcontents
\section{Random Voting Model: Description}
We describe a model of elections, designated as the Random Voting Model (RVM), in which $c_i$ number of candidates contest at $i$-th electoral unit with $n_i$ electors (voters). In this model, each elector from the $i$-th electoral unit casts their vote for $j$-th candidate with a probability $p_{ij}$. These probabilities are assigned as follows: for each candidate, a number between $0$ and $1$ is drawn uniformly at random, which is assigned as an unnormalized probability weight $w_{ij}$ to that candidate. The weights are subsequently normalized to get the probability $p_{ij}, j = 1, 2 \dots c_i$ of receiving the vote of an elector. This can be mathematically stated as
\begin{equation}
    w_{ij} \sim \mathcal{U}(0, 1) \quad \text{and} \quad p_{ij} = \frac{w_{ij}}{\sum_k w_{ik}}, \text{ with } j = 1, 2 \dots c_i,
\end{equation}
where $\mathcal{U}(0, 1)$ denotes a uniformly distributed random variable in $(0,1)$.

In an election, if there are $n_i$ electors (voters) in $i$-th electoral unit, each elector votes for candidate $j$ independently with probability $p_{ij}$. Every voter votes exactly once. The candidate receiving the most votes $v_{i, w}$ is declared the winner, and the candidate securing the next largest number of votes $v_{i, r}$ is the runner-up. The \emph{margin of victory} $M_i$ is then defined to be the vote difference between the winner and the runner-up: \emph{i.e.} $M_i = v_{i, w} - v_{i, w}$. The empirical election data we employ (from 34 countries) shows that the top three candidates, on average, account for nearly 87\% of all votes polled in an election.
Hence, as part of the model specification, we fix the number of candidates in each electoral unit to be three, i.e., 
$c_i = 3$ for all $i$.

The only input to this model is the raw turnout data, i.e., the number of voters (who actually voted) in each constituency. For the model simulation, we use the turnout data of real elections as the total number of voters in different constituencies. To understand how simulations are performed, consider this notional example: if a country has $N=100$ constituencies and data for five such elections is available. Then, the model is simulated on $500$ electoral units. The number of electors in each electoral unit is taken from the consolidated turnouts. Such a simulation of election is performed multiple times to get the average distributions for scaled margins $f(M / \langle M \rangle)$ and scaled specific margins $F(x)$.

\section{Computing the Distribution of Specific Margin $\mu = \frac{M}{T}$}
As done in the previous section, we consider the case where $3$ candidates are contesting in an election. The weight assigned for the $j$-th candidate of the $i$-th electoral unit is $w_{ij}$. These weights are drawn independently at random from a uniform distribution between $0$ and $1$.  The corresponding probability $p_{ij}$ of receiving votes is calculated by normalizing these weights. Hence, we have the following,
\begin{equation}
    w_{ij} \sim \mathcal{U}(0, 1) \text{ and } p_{ij} = \frac{w_{ij}}{\sum_{k=1}^3 w_{ik}}; \text{ with } j = 1, 2, 3.
\end{equation}
For the rest of the analysis, we focus on a single ($i$-th) electoral unit with voter turnout $T$ and drop the corresponding index $i$ for brevity. Hence,
\mathtoolsset{centercolon}
\begin{equation}
    w_{ij} := w_j \text{ and } p_{ij} := p_j.
\end{equation}
For large turnout $(T \gg 1)$, it is reasonable to assume the number of votes received by $j$-th candidate is proportional to their probability $p_j$, in particular, $v_j \approx p_j T$. Hence, for $T \gg 1$, the \emph{margin} can be approximated as 
\begin{equation}
M \approx (p_{max} - p_{2nd \: max})T,
\end{equation}
where $p_{max}$ and $p_{2nd \:max}$ correspond to the largest and the second largest probabilities assigned to the candidates. For example, if the probabilities $p_1, p_2,$ and $p_3$ assigned to the 3 candidates are $0.1, 0.6,$ and $0.3$, then $p_{max} = p_2 = 0.6$ and $p_{2nd \:max} = p_3 = 0.3$. The margin $M$ can also be written in terms of $w_j$ as the following:
\begin{center}
\begin{align}
    \nonumber M &\approx \left(\frac{w_{max}}{w_1 + w_2 + w_3} - \frac{w_{2nd\:max}}{w_1 + w_2 + w_3}\right)T,\\
    \nonumber & = \left(\frac{w_{(3)}}{w_{(1)} + w_{(2)} + w_{(3)}} - \frac{w_{(2)}}{w_{(1)} + w_{(2)} + w_{(3)}}\right)T,\\
    & = \left(\frac{w_{(3)} - w_{(2)}}{w_{(1)} + w_{(2)} + w_{(3)}}\right)T,
\end{align}

\end{center}
where $w_{(k)}$ is the $k$-th order statistics \cite{BarBalNag2008}. Hence, 
\begin{center}
\begin{align}
    \frac{M}{T} \approx \frac{w_{(3)} - w_{(2)}}{w_{(1)} + w_{(2)} + w_{(3)}}.
    \label{eq:S6}
\end{align}

\end{center}

Consider $n$ \emph{iid} random variables $\{X_1, X_2 \dots X_n\}$ drawn from a distribution $\rho(x)$. When arranged in ascending order, the random variable at the $k$-th spot is defined as the $k$-th order statistics. In particular, $n$-th and $1$-st order statistics correspond to the maximum and minimum of those $n$ random variables, respectively. The $k$-th order statistics of the random variable $X$ is denoted by $X_{(k)}$.

The joint probability density of all the order statistics of the above-mentioned $n$ random variables, $\mathbbm{P}\left(x_{(1)}, x_{(2)}, ... x_{(n)}\right)$, defined as the probability density that the random variable $X_{(k)}$ takes the value $x_{(k)}$ for $k \in \{ 1, 2, \dots, n\}$, is
\begin{equation}
    \mathbbm{P}\left(x_{(1)}, x_{(2)}, ... x_{(n)}\right) = n!\prod_{k=1}^{n}\rho\left(x_{(n)}\right).
\end{equation}
For our case, $n = 3$ and $\rho(x) = \mathcal{U}(0, 1)$. Hence we have,
\begin{center}
    \begin{align}
        \mathbbm{P}\left(w_{(1)}, w_{(2)}, w_{(3)}\right) = 3! = 6; \text{ with } 0<w_{(1)}<w_{(2)}<w_{(3)}<1,
    \end{align}
\end{center}
and $\mathbbm{P}\left(w_{(1)}, w_{(2)}, w_{(3)}\right) = 0$ otherwise, with the following normalization:
\begin{equation}
    \int_{0}^{1}dw_{(3)}\int_{0}^{w_{(3)}}dw_{(2)}\int_{0}^{w_{(2)}} 6 dw_{(1)} = 1.
\end{equation}
From the joint probability distribution of all the order statistics, we calculate the approximate probability density function of specific margin $ M / T = \mu$ from Eq.~\eqref{eq:S6} as follows, 
\begin{center}
    \begin{align}
        \nonumber P\left(\mu\right) & = 6 \nonumber \int_{0}^{1}dw_{(3)}\int_{0}^{w_{(3)}}dw_{(2)}\int_{0}^{w_{(2)}} \delta\left(\mu - \frac{w_{(3)}- w_{(2)}}{w_{(1)} + w_{(2)} + w_{(3)}}\right)dw_{(1)},\\
        & = 6 \int_{0}^{1}dw_{(3)}\int_{0}^{w_{(3)}} \frac{w_{(3)} - w_{(2)}}{\mu^2} \nonumber \mathbbm{1}_{0<\frac{w_{(3)} - \mu w_{(3)} - (1 + \mu)w_{(2)}}{\mu}<w_{(2)}} dw_{(2)},\\
        & = 6 \int_{0}^{1}dw_{(3)} \frac{(1 - \mu)(5 + 7\mu)w_{(3)}^2}{2(1 + \mu)^2(1 + 2\mu)^2}.\\
    \end{align}
\end{center}
Finally, after performing this integral, we get 
\begin{equation}
    P(\mu) = \frac{(1 - \mu)(5 + 7\mu)}{(1 + \mu)^2(1 + 2\mu)^2}.
\end{equation}
The distribution $P(\mu)$ does not depend on the turnout and is universal. Now, by a change of variable to scaled specific margin defined as $x = \mu / \langle \mu \rangle$, we obtain its distribution $F(x)$ to be
\begin{equation}
    F\left(x\right) = \langle \mu \rangle ~ P\left( x \langle \mu \rangle \right) =  \frac{\langle \mu \rangle(1 - x \langle \mu \rangle)(5 + 7x \langle \mu \rangle)}{(1 + x \langle \mu \rangle)^2(1 + 2x \langle \mu \rangle)^2}, 
\end{equation}
where $\langle \mu\rangle = \frac{1}{2}+\ln \left(\frac{9 \sqrt[4]{3}}{16}\right)$.

\section{Distribution of Margins and Their Tail Behaviors}
In this section, we obtain the distribution of margins $Q(M)$ for arbitrary turnout distribution $g(T)$, using the specific margin distribution $P(\mu)$. From the previous section, we have
\begin{equation}
    P(\mu) = \frac{(1 - \mu)(5 + 7\mu)}{(1 + \mu)^2(1 + 2\mu)^2}.
\end{equation}

Through a simple change of variable $(M = \mu T)$ we get,
\begin{equation}
    \mathcal{P}(M|T) = \frac{(1 - M / T)(5 + 7M /T)}{T(1 + M / T)^2(1 + 2M / T)^2}.
\end{equation}

For an arbitrary turnout distribution $g(T)$, we obtain the distribution of $M$ to be,
\begin{equation}
    Q(M) = \int_{M}^{\infty}g(T)\mathcal{P}(M |T) dT = \int_{M}^{\infty}g(T)\frac{(1 - M / T)(5 + 7M /T)}{T(1 + M / T)^2(1 + 2M / T)^2} dT.
\end{equation}

Again with $u = T / M$, the above integral transforms to,
\begin{equation}
    Q(M) = \int_{1}^{\infty}g(Mu)\frac{u(u - 1)(5u + 7)}{(1 + u)^2 (2 + u)^2}du.
    \label{eq:pm}
\end{equation}

We compute $Q(M)$ for different turnout distributions $g(T)$. In particular, we take $g(T)$ to be (A) exponential, (B) power law, and (C) Gaussian distributions as they have vastly different tail behaviors. 

\subsection{Exponential Turnout Distribution}
In this case $g(T) = \frac{1}{\tau}e^{-T / \tau}$, with $ \tau> 0$. Hence,
\begin{equation}
    Q(M) = \int_{1}^{\infty}\frac{1}{\tau}e^{-Mu / \tau} \frac{u(u - 1)(5u + 7)}{(1 + u)^2 (2 + u)^2}du,
\end{equation}
or,
\begin{equation}
    Q(M) = \frac{e^{-\frac{M}{\tau}}}{\tau^2} \left(4 e^{\frac{2 M}{\tau}} (\tau+M) \text{Ei}\left(-\frac{2 M}{\tau}\right)-9 e^{\frac{3 M}{\tau}} (\tau+2 M) \text{Ei}\left(-\frac{3 M}{\tau}\right)-4 \tau\right), 
\end{equation}

where $\text{Ei}(x) = \int_{-\infty}^{x}\frac{e^t}{t}dt$. At large margin limit $(M \rightarrow \infty)$, the asymptotic behavior of the distribution is the following (up to the leading order of $M$):
\begin{equation}
    Q(M)= \frac{\tau}{3M^2}e^{-M/\tau}.
\end{equation}

This suggests that in the large margin limit, both the margin and its corresponding turnout distribution have an exponential decay with the same rate.
\subsection{Power law Turnout Distribution}
In this case $g(T) = \frac{\alpha - 1}{T_{min} ^{1 -\alpha}} T ^ {-\alpha}$, with $\alpha > 1$ and $T>T_{min}$. Hence we have,
\begin{equation}
    Q(M) = \int_{1}^{\infty}\frac{\alpha - 1}{T_{min} ^{1 -\alpha}} (Mu) ^ {-\alpha} \frac{u(u - 1)(5u + 7)}{(1 + u)^2 (2 + u)^2}du,
\end{equation}
or,
\begin{equation}
    Q(M) = C(M)\frac{\alpha - 1}{T_{min} ^{1 -\alpha}} (M) ^ {-\alpha}, 
    \label{eq:powerlaw}
\end{equation}
where,
\begin{numcases}{C(M) = }
    I_1(\infty) - I_1(T_{min} / M) , \text{if } M\leq T_{min}\\
    I_1(\infty) - I_1(1), \text{otherwise,}
\end{numcases}
with,
\begin{equation}
    I_1(y) = \int \frac{y^{1 - \alpha}(y - 1)(5y + 7)}{(1 + y)^2 (2 + y)^2}dy,
\end{equation}
and,
\begin{numcases}{I_1(y)=}
     -\frac{4}{y+1}+\frac{9}{2 (y+2)}-\frac{1}{4} 7 \ln (y)+4 \ln (y+1)-\frac{9}{4} \ln (y+2), \text{if } \alpha = 2 \\
     \frac{y^{2-\alpha } \left(16 \, _2F_1(2,2-\alpha ;3-\alpha ;-y)-9 \, _2F_1\left(2,2-\alpha ;3-\alpha ;-\frac{y}{2}\right)\right)}{4 (\alpha -2)}, \text{otherwise,} \\
\end{numcases}
where ${}_{2}F_{1}(a,b;c;z)$ is a hypergeometric function \cite{abramowitz_stegun}, defined as,
\begin{align*}
    {\displaystyle {}_{2}F_{1}(a,b;c;z)  =\sum _{n=0}^{\infty }{\frac {(a)_{n}(b)_{n}}{(c)_{n}}}{\frac {z^{n}}{n!}}=1+{\frac {ab}{c}}{\frac {z}{1!}}+{\frac {a(a+1)b(b+1)}{c(c+1)}}{\frac {z^{2}}{2!}}+\cdots .}\\
\end{align*}

It is evident from Eq.~\eqref{eq:powerlaw} that for $M > T_{min}$, the margin distribution decays with a power law exponent $\alpha$, exactly the same as the turnout distribution.

\subsection{Gaussian Turnout Distribution}
In this case $g(T) = C_0 e^{-(T/T_0)^2}$, with $T>0$. Hence,
\begin{equation}
     Q(M) = \int_{1}^{\infty} C_0 e^{-(Mu/T_0)^2}\frac{u(u - 1)(5u + 7)}{(1 + u)^2 (2 + u)^2}du.
\end{equation}
At large margin limit $(M \rightarrow \infty)$, the asymptotic behavior of the distribution is the following (up to the leading order of $M$):
\begin{equation}
    Q(M) = \frac{C_0}{12}\left(\frac{T_0}{M}\right)^4 e^{-\left(M/ T_0\right)^2}, 
\end{equation}

and it has a Gaussian decay similar to the corresponding turnout distribution.\\

From the asymptotic analysis of the margin distributions for the three above-mentioned turnout distributions, we provide strong evidence that the tails of the margin distributions mimic that of the corresponding turnout distribution. For completeness, we also compute the margin distribution corresponding to a uniform turnout distribution which has a finite support (no tail behavior).

\subsection{Uniform Turnout Distribution}
In this case $g(T) = \frac{1}{b - a}$, when $T \in [a, b]$, otherwise  $g(T) = 0$. Hence,

\begin{numcases}{Q(M)= }
    \frac{1}{b - a}\int_{a/M}^{b/M}\frac{u(u - 1)(5u + 7)}{(1 + u)^2 (2 + u)^2}du , \text{if } M\leq a\\
    \frac{1}{b - a}\int_{1}^{b/M}\frac{u(u - 1)(5u + 7)}{(1 + u)^2 (2 + u)^2}du, \text{otherwise,}
\end{numcases}
or, 
\begin{numcases}{Q(M)= }
    \frac{1}{b - a} \left(I_2(b / M) - I_2(a / M)\right), \text{if } M\leq a\\
    \frac{1}{b - a}\left(I_2(b / M) - I_2(1)\right),  \text{if } a > M \geq b\\
    0,  \text{ otherwise,}
\end{numcases}
where, 
\begin{equation}
    I_2(y) = \int \frac{y(y - 1)(5y + 7)}{(1 + y)^2 (2 + y)^2}dy = -\frac{4}{y+1}+\frac{18}{y+2}-4 \ln (y+1)+9 \ln (y+2).
\end{equation}

\newpage

\color{black}
\section{RVM Simulations with Synthetic Turnout Distributions}

The RVM enables us to estimate the scaled margin distribution $f(M / \langle M \rangle)$ using only the raw turnout data, indicating that $f(M / \langle M \rangle)$ is driven by the details of the turnout distribution $g(T)$. To further quantify the effect of $g(T)$ on the scaled margin distribution $f(M / \langle M \rangle)$, we simulate elections using RVM, with turnouts drawn from vastly different synthetically generated distributions. In particular, to study the tail behaviors, we use the following four different turnout distributions:
\begin{enumerate}
    \item \textbf{Gaussian Turnout Distribution:} $g(T) = \frac{1}{\sigma\sqrt{2\pi}}\exp\left(-\frac{(T - \mu)^2}{2\sigma^2}\right), \text{ with } \mu = 50000$, $\sigma = 10000$ and $T > 0$.
    \item \textbf{Exponential Turnout Distribution:} $g(T) = \frac{1}{\tau} \exp{\left(-\frac{T}{\tau}\right)}, \text{ with } \tau = 50000$.
    \item \textbf{Power law Turnout Distribution:} $g(T) = \frac{\alpha - 1}{T_{min} ^{1 -\alpha}} T ^ {-\alpha}$, with $\alpha = 2$ and $T_{min} = 100$ (minimum possible turnout).
    \item \textbf{Uniform Turnout Distribution:} $T \sim \mathcal{U} (a, b)$, with $a = 100$ and $b = 100000$. $\mathcal{U}(a, b)$ denotes uniform distribution between the range $a$ and $b$.
\end{enumerate}

\begin{figure}[ht]
    \centering
    \includegraphics[width=0.95\textwidth]{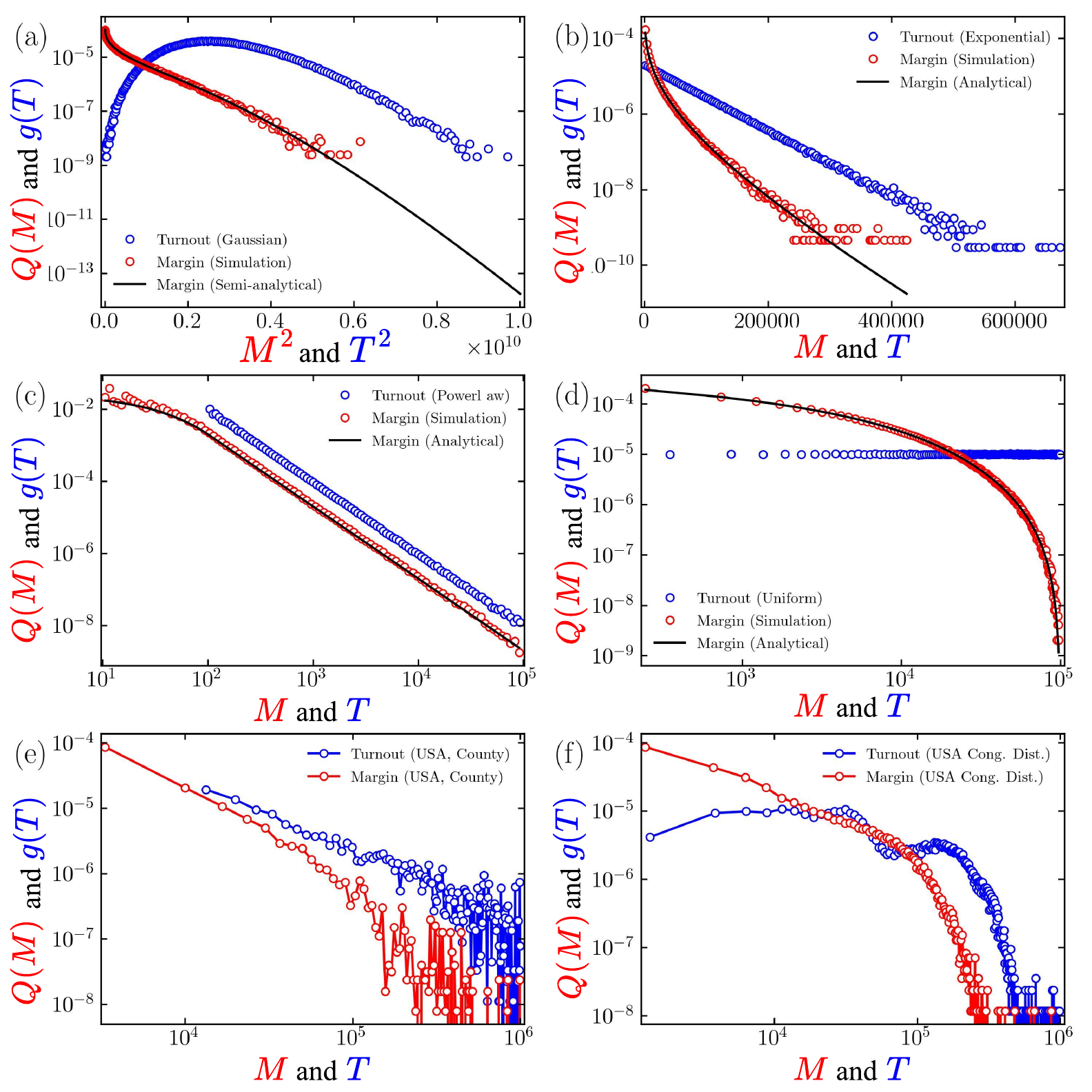}
    \caption{The margin distribution $Q(M)$ is plotted with the corresponding turnout distribution $g(T)$ to demonstrate that the tails of both these distributions are correlated. Panels (a), (b), (c), and (d) correspond to Gaussian, exponential, power law, and uniform turnout distributions, respectively. Blue open circles denote the turnout distributions. Red open circles denote the margin distribution computed through RVM simulations. Black solid lines correspond to the margin distribution computed using Eq.~\ref{eq:pm}. For exponential, power law, and uniform turnout distributions, the integration was analytically calculated, and for Gaussian turnout distribution, it was evaluated numerically. Panels (e) and (f) depict the margin and turnout distribution for the county-level and congressional district-level election data of the USA, respectively.}
    \label{fig_sup_1}
\end{figure}

Each of the RVM simulations was performed on $10^6$ electoral units, with turnouts (rounded down to the nearest integer) drawn from one of these three distributions. The simulation demonstrates that the tail of the margin distribution mimics the turnout distribution's tail.  This is evident in Fig.~\ref{fig_sup_1}(a), (b), and (c). The tail of the margin distribution (Fig.~\ref{fig_sup_1} (c)) corresponding to power law turnouts decays with the same power law exponent.  In the simulation with Gaussian turnout distribution, we find the tail of the margin distribution also has a Gaussian falloff (Fig.~\ref{fig_sup_1} (a)). Similarly, the margin distribution corresponding to exponential turnouts has an exponential tail (Fig.~\ref{fig_sup_1} (b)). As the probability density function of uniform turnout distribution and corresponding margin distribution have finite supports, their tails can not be properly defined. We find a sharp cutoff in the corresponding margin distribution. The analytical (semi-analytical for Gaussian turnout) predictions for the margin distributions (shown as black lines in Fig.~\ref{fig_sup_1}) corresponding to all four aforementioned turnout distributions are in excellent agreement with the RVM simulation. In empirical county-level election data of the United States, the heavy-tailed decay of the turnout distribution is reflected in the corresponding margin distribution (Fig.~\ref{fig_sup_1}(e)). In Fig.~\ref{fig_sup_1} (f), we see a similar decay trend in both margin and turnout distribution, which correspond to congressional district-level election data of the USA. We obtain the scaled margin distribution $f(M / \langle M \rangle)$ by scaling $Q(M)$ by its mean; hence, both $Q(M)$ and $f(M / \langle M \rangle)$ have similar decay and are strongly related to the corresponding turnout distribution $g(T)$.

\vspace{1em}
\noindent \textbf{Simulation details of the universality result:} We study the scaled specific margin distribution $F(x)$ by simulating elections using RVM for the following three turnout distributions:
\begin{enumerate}
    \item \textbf{Gaussian Turnout Distribution:} $g(T) = \frac{1}{\sigma\sqrt{2\pi}}\exp\left(-\frac{(T - \mu)^2}{2\sigma^2}\right), \text{ with } \mu = 50000$, $\sigma = 10000$ and $T > 0$.
    \item \textbf{Uniform Turnout Distribution:} $T \sim \mathcal{U} (a, b)$, with $a = 100$ and $b = 100000$. $\mathcal{U}(a, b)$ denotes uniform distribution between the range $a$ and $b$.
        \item \textbf{Power law Turnout Distribution:} $g(T) = \frac{\alpha - 1}{T_{min} ^{1 -\alpha}} T ^ {-\alpha}$, with $\alpha = 2$ and $T_{min} = 100$ (minimum possible turnout).
\end{enumerate}
Turnouts drawn from these distributions are rounded down to the nearest integers. Simulations performed with a large number of electoral units $(10^6)$ lead to a perfect collapse in the scaled specific margin distributions $F(x)$, which is in remarkable agreement with the theoretically predicted distribution, as shown in Fig.~~3(a) of the letter. When simulations are performed with realistic numbers of electoral units as found in the empirical data, the corresponding scaled distributions of $\mu$ show similar fluctuations around the universal curve, as found in the empirical distributions of individual countries. To ensure realistic statistics, the number of electoral units $N$ chosen for each simulation is the consolidated number of electoral units for each of the $32$ countries. Once the number of electoral units is fixed, we randomly choose one of the three distributions mentioned above. Further $N$ turnouts are drawn independently from that distribution, and the RVM simulation is performed on those turnouts.

\newpage

\newpage

\section{Scaled Margin Distributions for Different $p_{ij}$ Distributions} 
\begin{figure}[h!]
    \centering
    \includegraphics[width=1\textwidth]{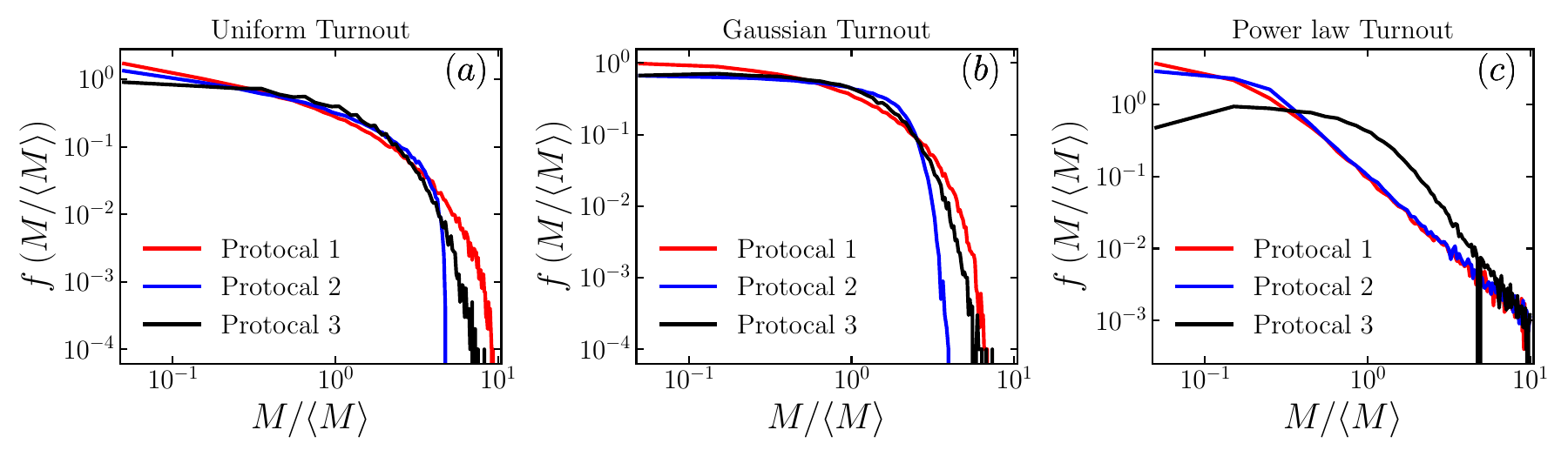}
    \caption{Prediction of scaled margin distribution for three different protocols of choosing $p_i$, the probability of receiving votes. Panels (a), (b), and (c) are for uniform, Gaussian, and Power law turnout distributions, respectively.}
    \label{fig_supp_3}
\end{figure}

To investigate the effect of the distribution of $p_{ij}$ on the prediction of scaled margin distribution, we simulated RVM using the following three protocols for choosing $p_{ij}$.
\begin{enumerate}
    \item \textbf{Protocol 1:} $w_{ij} \sim \mathcal{U}(0, 1) \text{ and } p_{ij} = \frac{w_{ij}}{\sum_{k=1}^3 w_{ik}}; \text{ with } j = 1, 2, 3$.
    \item \textbf{Protocol 2:} $w_{i1} \sim \mathcal{U}(0, 1), w_{i2} \sim \mathcal{U}(0, 1 - w_{i1}), w_{i3} = 1 - w_{i1} - w_{i2} \text{ and } p_{ij} = \frac{w_{ij}}{\sum_{k=1}^3 w_{ik}} = w_{ij}; \text{ with } j = 1, 2, 3$.
    \item \textbf{Protocol 3:} $w_{ij} = p_{ij} = \frac{1}{3}, \text{ with } j = 1, 2, 3$.
\end{enumerate}
In Fig.~\ref{fig_supp_3}, we demonstrate the differences in the prediction of scaled margin distributions for synthetically generated turnout distributions when the three aforementioned protocols are used. Panel (a) shows that, for turnouts drawn from a uniform distribution, the prediction using protocols 1 and 3 are similar, while protocol 2 produces a scaled margin distribution that decays faster. In panel (b), we see similar results for turnout drawn from a Gaussian distribution. However, panel (c) depicts that, for turnouts drawn from a power law distribution, the predictions using protocols 1 and 2 are almost identical, while protocol 3 produces a vastly different scaled margin distribution.

\newpage

\section{Data Collection and Cleaning}
In this work, we use empirical election data from 34 countries. Of these, data from 32 countries are used to establish the universality result, and data from two countries illustrate pronounced cases of deviations from universality.\\
 
\emph{Data collection}-- We collect constituency-level data of the lower chamber of the Legislative elections for 180 countries and territories across the world from the Constituency-Level Election Archive (CLEA) website \cite{clea}. Polling booth level data for India and Canada is collected from the websites of Election Commission \cite{india_data, canada_data} of the respective countries, semi-automatically using a combination of Python libraries. We collect county-level data from MIT Election Data + Science Lab \cite{DVN/VOQCHQ_2018} for the USA. While constituency-level data is available for many countries, polling booth-level data is available in the public domain only for a few countries.\\

\emph{Data cleaning}--While constituency-level data collected from the CLEA website was in tabular format, the polling booth-level data was found in different formats, ranging from tabular to machine-generated and scanned PDFs. We clean the data using a combination of Python libraries. Our analysis was performed on the election data of each country, which was consolidated over several elections. To ensure a reasonable level of confidence in the statistical analysis, we have ignored data from countries with less than 400 data points. By this criteria, we could use the data from $34$ out of $180$ countries, all of which have more than $400$ data points. The threshold of $400$ data points allows us to demonstrate universality, along with flagging possible electoral misconduct in Ethiopia and Belarus, while maintaining good statistics.\\ 

In this analysis, we discard those rare cases when the turnout is zero, or the number of contesting candidates is less than two. To avoid discrepancies, we consider the sum of valid votes received by all the candidates (in an electoral unit) as the turnout for the election in that unit. Some important summary statistics of the election data for the $34$ countries used for analysis in this work are given in Table \ref{table}.
\newpage

\begin{table}[h]
\centering
\begin{tabular}{|c|c|c|c|c|c|c|c|}
\hline
Country & Time span & Number & Scale    & Mean turnout & Mean margin  & Number \\ 
 &  & of  &  &  &  & of electoral \\ 
 &  & elections  &  &  &  & units \\ 
 &  &  &  &  &  & (consolidated) \\ \hline
Australia & 1901-2016 & 37 & Constituency & $7.37\times 10^{4}$ & $1.31\times 10^{4}$ & 1740\\ \hline
Bangladesh & 1973-2008 & 4 & Constituency & $1.57\times 10^{5}$ & $3.15\times 10^{4}$ & 1188\\ \hline
Belarus & 2004-2019 & 5 & Constituency & $4.83\times 10^{4}$ & $2.61\times 10^{4}$ & 441\\ \hline
Canada & 1867-2019 & 43 & Constituency & $2.76\times 10^{4}$ & $5.50\times 10^{3}$ & 10662\\ \hline
Canada & 2004-2021 & 7 & Polling Booth & $5.56\times 10^{2}$ & $1.35\times 10^{2}$ & 489919\\ \hline
Chile & 1945-2017 & 7 & Constituency & $1.07\times 10^{5}$ & $1.05\times 10^{4}$ & 420\\ \hline
Denmark & 1849-2019 & 30 & Constituency & $2.70\times 10^{3}$ & $4.64\times 10^{2}$ & 2178\\ \hline
Ethiopia & 2010-2010 & 1 & Constituency & $4.95\times 10^{4}$ & $4.18\times 10^{4}$ & 492\\ \hline
France & 1973-2017 & 3 & Constituency & $7.88\times 10^{4}$ & $1.10\times 10^{4}$ & 1712\\ \hline
Germany & 1871-2017 & 19 & Constituency & $1.37\times 10^{5}$ & $2.26\times 10^{4}$ & 5108\\ \hline
Ghana & 1992-2016 & 6 & Constituency & $3.75\times 10^{4}$ & $9.88\times 10^{3}$ & 1410\\ \hline
Hungary & 1990-2018 & 6 & Constituency & $5.32\times 10^{4}$ & $8.57\times 10^{3}$ & 936\\ \hline
India & 1951-2019 & 18 & Constituency & $5.69\times 10^{5}$ & $8.33\times 10^{4}$ & 8389\\ \hline
India & 2004-2019 & 4 & Polling Booth & $5.82\times 10^{2}$ & $1.89\times 10^{2}$ & 752786\\ \hline
Japan & 1947-2017 & 26 & Constituency & $2.88\times 10^{5}$ & $2.35\times 10^{4}$ & 4603\\ \hline
Kenya & 1961-2013 & 2 & Constituency & $3.72\times 10^{4}$ & $1.19\times 10^{4}$ & 417\\ \hline
Korea & 1948-2012 & 13 & Constituency & $6.17\times 10^{4}$ & $1.01\times 10^{4}$ & 2258\\ \hline
Lithuania & 1992-2020 & 8 & Constituency & $3.24\times 10^{4}$ & $3.98\times 10^{3}$ & 570\\ \hline
Malawi & 1994-2019 & 4 & Constituency & $2.31\times 10^{4}$ & $6.29\times 10^{3}$ & 755\\ \hline
Malaysia & 1959-2018 & 13 & Constituency & $3.41\times 10^{4}$ & $8.90\times 10^{3}$ & 2199\\ \hline
Myanmar & 2010-2015 & 2 & Constituency & $6.76\times 10^{4}$ & $2.32\times 10^{4}$ & 634\\ \hline
New Zealand & 1943-2020 & 9 & Constituency & $3.04\times 10^{4}$ & $6.94\times 10^{3}$ & 637\\ \hline
Nigeria & 2003-2019 & 2 & Constituency & $7.75\times 10^{4}$ & $2.20\times 10^{4}$ & 710\\ \hline
Pakistan & 1988-2013 & 3 & Constituency & $1.28\times 10^{5}$ & $2.45\times 10^{4}$ & 683\\ \hline
Papua New Guinea & 1972-2017 & 8 & Constituency & $5.07\times 10^{4}$ & $5.66\times 10^{3}$ & 841\\ \hline
Philippines & 1946-2013 & 17 & Constituency & $1.83\times 10^{5}$ & $2.63\times 10^{4}$ & 2525\\ \hline
Solomon Islands & 1967-2019 & 14 & Constituency & $3.67\times 10^{3}$ & $4.37\times 10^{2}$ & 543\\ \hline
Taiwan & 1986-2020 & 11 & Constituency & $2.33\times 10^{5}$ & $1.98\times 10^{4}$ & 482\\ \hline
Tanzania & 2005-2020 & 2 & Constituency & $5.37\times 10^{4}$ & $2.01\times 10^{4}$ & 492\\ \hline
Thailand & 1969-2011 & 12 & Constituency & $1.86\times 10^{5}$ & $1.46\times 10^{4}$ & 2263\\ \hline
Trinidad and Tobago & 1925-2020 & 13 & Constituency & $1.53\times 10^{4}$ & $5.12\times 10^{3}$ & 411\\ \hline
Uganda & 2006-2021 & 4 & Constituency & $4.45\times 10^{4}$ & $1.08\times 10^{4}$ & 1430\\ \hline
UK & 1832-2019 & 46 & Constituency & $3.43\times 10^{4}$ & $6.30\times 10^{3}$ & 23105\\ \hline
Ukraine & 1998-2019 & 5 & Constituency & $8.89\times 10^{4}$ & $1.67\times 10^{4}$ & 1072\\ \hline
United States & 1788-2020 & 167 & Congressional District & $1.14\times 10^{5}$ & $2.96\times 10^{4}$ & 33946\\ \hline
United States & 2000-2020 & 6 & County & $1.78\times 10^{5}$ & $2.00\times 10^{4}$ & 18905\\ \hline
Zimbabwe & 2005-2018 & 4 & Constituency & $1.77\times 10^{4}$ & $6.55\times 10^{3}$ & 743\\ \hline
\end{tabular}
\caption{Typical values of Margin and turnouts at different scales for different countries. The available data for the mentioned time spans were consolidated for each country and used to calculate the mean turnout and mean margin. The consolidated number of electoral units (in the last column) is calculated by adding the number of valid electoral units for all the elections that happened in the mentioned time span. The data for an electoral unit is considered to be valid if (a) a list of votes received by all the candidates is available, (b) at least two candidates are contesting, and (c) the turnout is non-zero. For example, in polling booth level data for India, lists of votes for all the polling booths are not always available. We could obtain valid data for $752786$ polling booths from the four elections held during the time span of $2004-2019$ for which the above-mentioned conditions were met. Only national-level elections are considered in this dataset.}
\label{table}
\end{table}

\newpage
\section{Figures Containing $f\left(M / \langle M\rangle\right)$ for $32$ Countries}
\begin{figure}[h!]
    \includegraphics[width=0.85\textwidth]{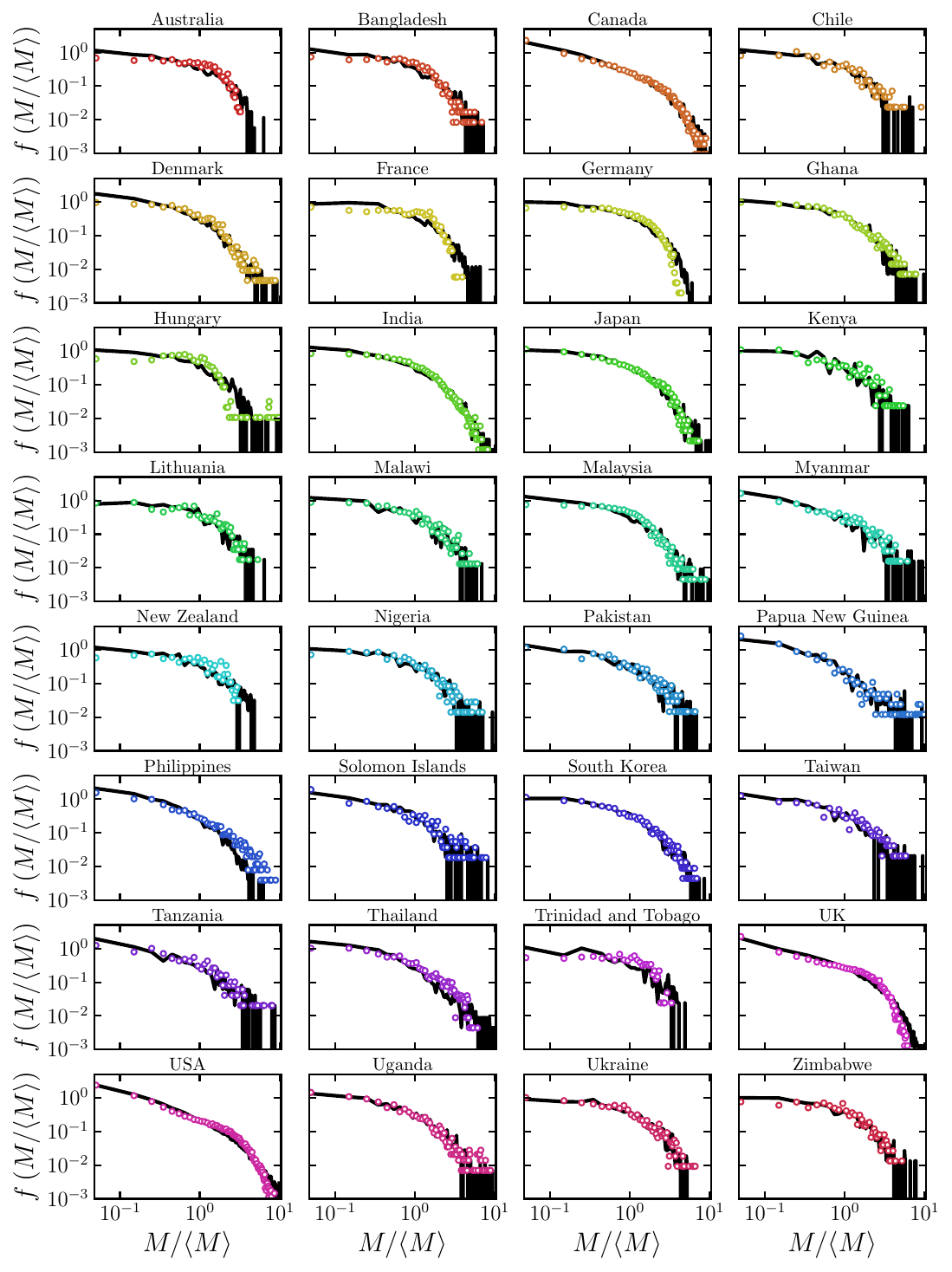}
    \caption{The empirical distribution of the scaled margins (colored open circles), along with RVM model prediction (black solid lines) for $32$ countries.}
    \label{fig_sup_4}
\end{figure}
\newpage

\section{Figures Containing $F\left(x\right)$ for $32$ Countries}
\begin{figure}[h!]
    \includegraphics[width=0.85\textwidth]{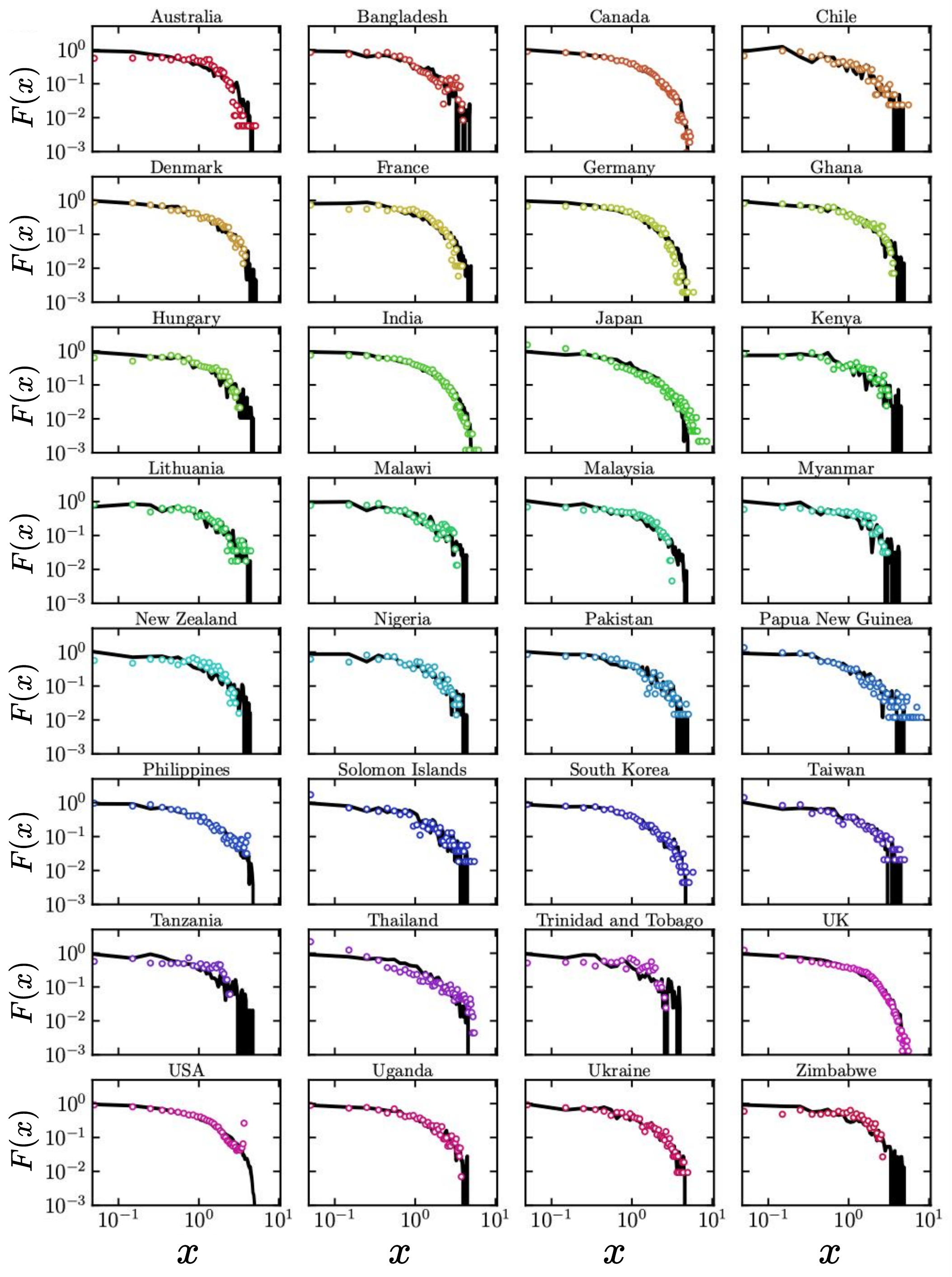}
    \caption{The empirical distribution of the scaled specific margin (colored open circle), along with RVM model prediction (black solid line) for $32$ countries.}
    \label{fig_sup_5}
\end{figure}

\newpage

\section{Scaling of $\langle M \rangle$ and $\langle \mu \rangle$ vs $T$}
\begin{figure}[ht]
    \centering
    \includegraphics[width=1\linewidth]{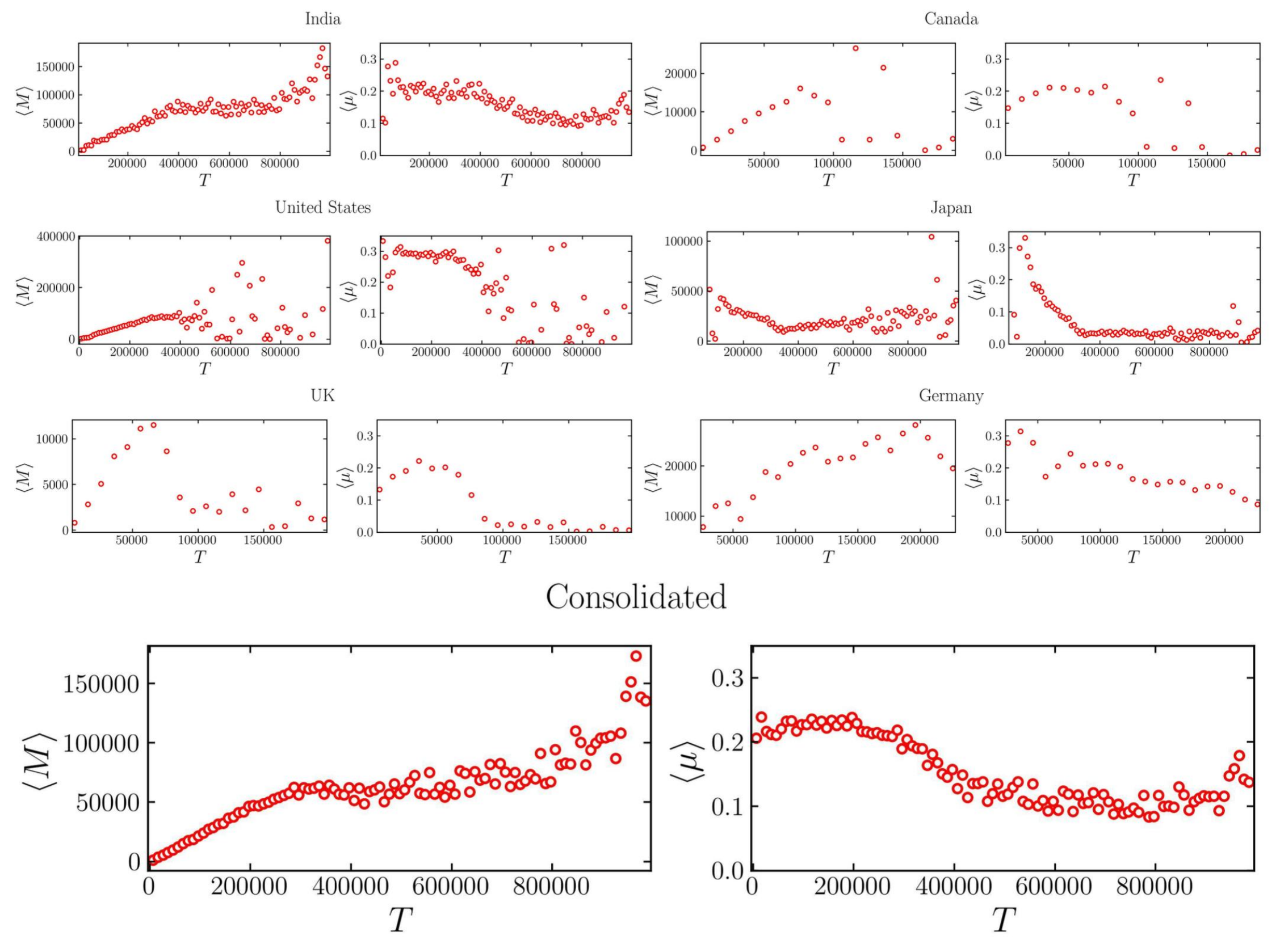}
    \caption{$\langle M \rangle$ vs $T$ and $\langle \mu \rangle$ vs $T$ for India, Canada, the United States, Japan, the UK, and Germany. The bottom-most panels display plots of the consolidated election data, combining data from 32 countries.}
    \label{fig_supp_2}
\end{figure}

At a large turnout limit $(T \gg 1)$, the distributions of $\mu$ and $M$ produced by RVM are the following, 
\begin{align*}
    P(\mu) =  \frac{(1 - \mu)(5 + 7\mu)}{(1 + \mu)^2(1 + 2\mu)^2},\\
    \mathcal{P}(M|T) = \frac{(1 - M / T)(5 + 7M /T)}{T(1 + M / T)^2(1 + 2M / T)^2}.
\end{align*}
From this, we find $\langle \mu \rangle = \frac{1}{2}+\ln \left(\frac{9 \sqrt[4]{3}}{16}\right)$ and $\langle M\rangle = T\left(\frac{1}{2}+\ln \left(\frac{9 \sqrt[4]{3}}{16}\right)\right)$. We investigate if such linear scaling of $\langle M \rangle$ with $T$ exists in empirical data. As shown in Fig.~\ref{fig_supp_2}, for some countries (India, Canada, the United States, and the UK), there is a region of linearity in the $\langle M \rangle$ vs $T$ plots. Correspondingly, $\langle \mu  \rangle$ is constant in those regions. Japan and Germany pose as counterexamples to this linearity hypothesis, and developing a better understanding of this scaling provides a rich avenue for further research. 

\end{document}